\documentclass[preprint,12pt]{elsarticle}

\usepackage{graphicx}
\usepackage{textgreek}
\usepackage{amssymb}

\usepackage{lineno}

\usepackage[hyphens]{url}

\journal{Computational and Structural Biotechnology Journal }

\begin{document}

\begin{frontmatter}


\title{\bf\large FHIRChain: Applying Blockchain to Securely \\ and Scalably Share Clinical Data}



\author[eecs]{Peng Zhang}
\ead{peng.zhang@vanderbilt.edu}

\author[eecs]{Jules White}
\ead{jules.white@vanderbilt.edu}

\author[eecs]{Douglas C. Schmidt}
\ead{d.schmidt@vanderbilt.edu}

\author[gl]{Gunther Lenz}
\ead{gunther.lenz@varian.com}

\author[tr]{S. Trent Rosenbloom}
\ead{trent.rosenbloom@vanderbilt.edu}

\address[eecs]{Department of Electrical Engineering and Computer Science, Vanderbilt University, Nashville, Tennessee, USA}

\address[gl]{Varian Medical Systems, Palo Alto, California, USA}

\address[tr]{Department of Biomedical Informatics, Vanderbilt University Medical Center, Nashville, Tennessee, USA}

\begin{abstract}
Secure and scalable data sharing is essential for collaborative clinical decision making.  Conventional clinical data efforts are often siloed, however, which creates barriers to efficient information exchange and impedes effective treatment decision made for patients.  This paper provides four contributions to the study of applying blockchain technology to clinical data sharing in the context of technical requirements defined in the ``Shared Nationwide Interoperability Roadmap" from the \textit{Office of the National Coordinator for Health Information Technology} (ONC).  First, we analyze the ONC requirements and their implications for blockchain-based systems.  Second, we present FHIRChain, which is a blockchain-based architecture designed to meet ONC requirements by encapsulating the HL7 \textit{Fast Healthcare Interoperability Resources} (FHIR) standard for shared clinical data. Third, we demonstrate a FHIRChain-based decentralized app using digital health identities to authenticate participants in a case study of collaborative decision making for remote cancer care.  Fourth, we highlight key lessons learned from our case study. 	
\end{abstract}


\end{frontmatter}


\section{Introduction}
\label{intro}

\textbf{The importance of data sharing in collaborative decision making}. Secure and scalable data sharing is essential to provide effective collaborative treatment and care decisions for patients. Patients visit many different care providers' offices during their lifetime. These providers should be able to exchange health information about their patients in a timely and privacy-sensitive manner to ensure they have the most up-to-date knowledge about patient health conditions. 

As another example, in telemedicine practice~\cite{berman2005technology}---where patients are remotely diagnosed and treated---the ability to exchange data securely and scalably is of paramount importance.  Data sharing helps improve diagnostic accuracy~\cite{castaneda2015clinical} by gathering confirmations or recommendations from a group of medical experts, as well as preventing inadequacies~\cite{singh2013types} and errors in treatment plan and medication~\cite{kaushal2003effects,schiff2009diagnostic}. Likewise, aggregated intelligence and insights~\cite{taichman2016sharing, warren2016strengthening, geifman2015opening} helps clinicians understand patient needs and in turn apply more effective in-person and remote treatments. 

Data sharing is also essential in cancer care, where groups of physicians with different specialties form tumor boards. These boards meet on a regular basis to analyze cancer cases, exchange knowledge, and collaboratively create effective treatment and care plans for each patient~\cite{Gross1987}.  Regional virtual tumor boards are also being implemented via telemedicine~\cite{ricke2000telemedicine,marshall2014implementation} for institutions that lack inter-specialty cancer care due to limited oncology expertise and resources~\cite{levit2010ensuring}.  

\textbf{Administrative support for coordinating health IT efforts}. The Office of the National Coordinator for Health Information Technology (ONC) is a division of the Office of the Secretary within the United States Department of Health and Human Services.  ONC is the principal federal entity to oversee and coordinate health IT efforts, including the development of interoperable, privacy-preserving, and secure nationwide health information systems and the promotion of widespread, meaningful use of health IT to improve healthcare. 

\textbf{Data sharing barriers to collaborative decision making}. In practice, many barriers exist in the technical infrastructure of health IT systems today that impede the secure and scalable data sharing across institutions, thereby limiting support for collaborative clinical decision making.  Examples of such barriers include the following:
\begin{itemize}
\item \textbf{Security and privacy concerns.} Despite the need for data sharing, concerns remain regarding protection of patient identity and confidentiality~\cite{terry2009medical}.  For instance, virtual medical interactions may increase the risk of clinical data breaches due to electronic transmission of data without highly secure infrastructures in place, which can result in severe financial and legal consequences~\cite{Downey2013}.  Likewise, medical identity theft may occur more frequently, especially in telemedicine~\cite{terry2009medical}, where virtual (\textit{i.e.}, networked) interactions are replacing face-to-face interactions between providers and patients.
\item \textbf{Lack of trust relationships between healthcare entities}.  Trust relationships between healthcare entities~\cite{Hripcsak2014} (\textit{e.g.}, care providers and/or healthcare institutions)  are an important precondition to digital communications~\cite{hartvigsen2007challenges} and data sharing in the absence of custody over shared data.  Larger healthcare facilities (such as enterprise hospital systems) may be networked~\cite{maheu2002health}, but communications between private or smaller practices may not be established.
\item \textbf{Scalability concerns}.  Large-scale datasets may be hard to transmit electronically due to restrictive firewall settings or limitations in bandwidth (which is still common in rural areas~\cite{larose2011impact}).  Lack of scalability can also impact overall system response time and data transaction speed~\cite{bondi2000characteristics}. 
\item \textbf{Lack of interoperable data standards enforcement}.  Without the enforcement of existing interoperable data standards (such as HL7's \textit{Fast Healthcare Interoperability Resources} (FHIR)\cite{bender2013hl7} for shared data), health data can vary in formats and structures that are hard to interpret and integrate into other systems~\cite{Richesson2007}.  
\end{itemize}

What is needed, therefore, is a standards-based architecture that can integrate with existing health IT systems (and related mobile apps) to enable secure and scalable clinical data sharing for improving continuous, collaborative decision support. 

\textbf{Research focus and contributions $\rightarrow$ Architectural considerations for secure and scalable blockchain-based clinical data sharing systems}. Blockchain technologies have recently been touted~\cite{das2017does,mettler2016blockchain,azaria2016medrec} as a technical infrastructure to support clinical data sharing that promotes care coordination. A key property of blockchains is their support for "trustless disintermediation." This property enables multiple parties who do not fully trust each other to exchange digital assets (such as the Bitcoin cryptocurrency~\cite{nakamoto2008bitcoin}), while still protecting their sensitive, personal data from each other. 

Our prior work~\cite{zhang2017metric} provided evaluation recommendations for blockchain-based health IT solutions on a high-level, focusing on common software patterns~\cite{zhang2017applying} that can be applied to improve the design of blockchain-based health apps. This paper examines previously unexplored research topics related to alleviating the data sharing barriers described above, namely: \textit{what are the architectural consideration associated with properly leveraging blockchain technologies to securely and scalably share healthcare data for improving collaborative clinical decision support}? 

This paper provides the following contributions to using blockchain technologies in clinical data sharing to improve collaborative decision support: 
\begin{itemize}
\item We summarize key technical requirements defined in the ``Shared Nationwide Interoperability Roadmap"~\cite{desalvo2015connecting} drafted by the \textit{Office of the National Coordinator for Health Information Technology} (ONC) for creating an interoperable health IT system and analyze the implications for blockchain-based system design. 
\item We present the structure and funcationality of a blockchain-based architecture called FHIRChain that meets the ONC technical requirements for sharing clinical data between distributed providers.  FHIRChain uses HL7's FHIR data elements (which have uniquely identifying tags) in conjunction with a token-based design to exchange data resources in a decentralized and verifiable manner without requiring duplicated efforts of uploading data to a centralized repository.  
\item We demonstrate a FHIRChain-based \textit{decentralized app} (DApp) that uses digital health identities to more readily authenticate participants and manage data access authorizations in a case study of clinical data sharing in remote cancer care. This DApp enables users to share specific and structured pieces of information (rather than an entire document), thereby increasing the readability of data and flexibility of sharing options. 
\item We highlight key lessons learned from our case study and discuss how our FHIRChain-based DApp can be further extended to support other technical requirements for improving advanced healthcare interoperability issues, such as coordinating other stakeholders (\textit{e.g.}, insurance companies and pharmacies) across the industry and providing patients with direct and secure access to their own medical records. We also explore the data exchange issues that blockchains cannot yet address effectively, including semantic interoperability, healthcare malpractice, and unethical use of the data, which remain as future research problems in this space.
\end{itemize}

\textbf{Paper Organization.}  The remainder of this paper is organized as follows: Section~\ref{background} provides an overview of blockchain technologies and the Ethereum platform, which is an open-source blockchain implementation that supports the development of DApps via ``smart contracts;'' Section~\ref{related} surveys different  blockchain-based research approaches in the healthcare domain and compares our research on FHIRChain with related work; Section~\ref{requirements} summarizes ONC's key technical requirements for sharing clinical data and analyzes their implications for blockchain-based designs; Section~\ref{fhirchain} describes how the blockchain-based architecture of FHIRChain is designed to meet ONC requirements and motivates why we made certain architectural decisions; Section~\ref{results} analyzes the benefits and limitations of a case study that applied a FHIRChain-based DApp to provide collaborative clinical decision support; and Section~\ref{conclusion} presents concluding remarks and outlines our key lessons learned and future work on extending the FHIRChain architecture described in this paper.

\section{Overview of Blockchain }
\label{background}

\begin{figure}[htpb]
\centering
\includegraphics[width=0.8\columnwidth]{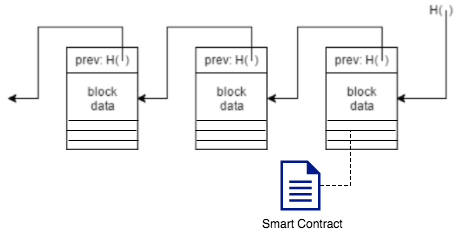}
\caption{The Blockchain Structure: a Continuously Growing and Immutable List of Ordered and Validated Transactions}
\label{fig:blockchain}
\end{figure}

The most popular application of blockchain is the Bitcoin blockchain~\cite{nakamoto2008bitcoin}, which is a public distributed ledger designed to support financial transactions via the Bitcoin cryptocurrency. This blockchain operates in a peer-to-peer fashion with all transactions distributed to each network maintainer node (called a ``miner'') for verification and admittance onto the blockchain. These miners validate available transactions and group them into blocks, as shown in Figure~\ref{fig:blockchain}. 
Miners then compete in solving a computationally expensive cryptographic puzzle, known as ``proof-of-work,'' where the first miner to solve this puzzle receiving a reward (\textit{i.e.}, an amount of Bitcoin) and appending their block of validated transactions to the blockchain sequence.  

The Bitcoin blockchain uses the “proof-of-work” process outlined above to achieve consensus by 
\begin{itemize}
\item incentivizing miners to contribute powerful hardware and electricity to the network with small amounts of cryptocurrency as rewards and 
\item discouraging rogue actors from attempting to manipulate or maliciously control the system. 
\end{itemize}
After a block is added to the blockchain, its transaction history is secured from tampering via cryptography. 

The Bitcoin blockchain is the most widely deployed example of this distributed ledger technology. In recent years, however, other types of blockchain technologies have emerged.  For example, the Ethereum blockchain~\cite{buterin2013ethereum} provides a more generalized framework via ``smart contracts"~\cite{johnston2014general} that allow programs to run on the blockchain and store/retrieve information. 

Smart contracts enable code to execute autonomously when certain conditions are met,  as shown in Figure~\ref{fig:blockchain}. They can also store information as internal state variables and define custom functions to manipulate or update this state. Operations in smart contracts are published as transactions and thus occur in a globally sequential order.  These operations are deterministic and verifiable by miners in the Ethereum blockchain to ensure their validity.

The mechanisms described above make a blockchain decentralized and immutable, thereby removing the need for a trusted central authority. These properties make blockchain technologies attractive to certain communities of health IT researchers and practitioners as means to improve clinical communications while protecting the privacy of healthcare participants. The remainder of this paper examines how to effectively leverage blockchains for securely and scalably sharing clinical data that enables collaborative decision support.

\section{Related Work Summary and Comparison}
\label{related}

Due to the growing interest in using distribute ledger technologies for health IT systems, related work has explored various blockchain-based design considerations and prototypes. This section summarizes this related work and compares it with our research on FHIRChain and DApps that provide collaborative clinical decision support for remote patients.



\subsection{Conceptual Blockchain-Based Design Considerations}
Krawiec et al.~\cite{krawiec2016blockchain} presented several existing pain points in current health information exchange systems and the corresponding opportunities provided by blockchain technologies. They also discussed how blockchain can be leveraged in the health IT systems so that patients, health providers, and/or health organisations can collaborate. Nichol et al~\cite{Nichol2016} presented an analysis that assembles concepts in blockchain-related technologies and speculates on how blockchain can be used to solve common interoperability problems facing healthcare.  

A team at IBM~\cite{Team2016} took a broader approach by highlighting the challenges in the healthcare industry and providing concrete use cases to showcase potential applications of blockchain technologies.  Our prior work also provided software design recommendations for creating general blockchain-based health IT systems~\cite{zhang2017applying} and proposed assessment metrics for blockchain-based health systems~\cite{zhang2017metric}, which include a subset of the technical requirements defined in the ONC roadmap.  This prior work of ours focused on providing more general or high-level recommendations for developers creating blockchain-based health IT systems.

The review paper by Kuo et al.~\cite{kuo2017blockchain} presented several blockchain applications in healthcare, such as improved medical record management and advanced healthcare data ledger, and their benefits for each described application. They then analyzed key challenges associated with using blockchain technology for healthcare, including issues like confidentiality, scalability, and treat of a 51\% attack on the blockchain network. According to the authors, some example implementation techniques that may mitigate the challenges are (1) encryption of sensitive data or dissemination of only meta data and storing sensitive data off-chain to protect confidentiality, (2) keeping only partial, ongoing verified transactions on-chain rather than the entire transaction history to increase scalability of the blockchain network, and (3) the adoption of a virtual private network or HIPAA-compliant components to prevent the 51\% attack.

\subsection{Blockchain Prototype Designs}
Ekblaw et al.~\cite{Ekblaw2016} created a decentralized record management platform that enables patients to access their medical history across multiple providers. This platform used a so-called ``permissioned'' blockchain (which is only accessible by authorized users, rather than one that is open to the public) to manage authentication, data sharing, and other security properties in the medical domain. Their blockchain design integrated with existing provider data storage to enable interoperability by curating a representation of patient medical records. Medical researchers were incentivised to contribute to mining of the blockchain by collecting aggregated metadata as mining rewards. 

Peterson et al.~\cite{Peterson2016} presented a healthcare blockchain also considers the integration with FHIR standards. They proposed a merkle-tree based blockchain system that introduces "Proof of Interoperability" as the consensus mechanism during block mining. Proof of interoperability is based on conformance to the FHIR protocol, meaning that miners must verify the clinical messages sent to their blockchain to ensure they are interoperable with known structural and semantic standards.

Dubovitskaya et al.~\cite{dubovitskaya2017secure} also proposed a permissioned blockchain framework on managing and sharing medical records for cancer patient care. Their design employed a membership service to authenticate registered users using a username/password scheme. Patient identity was created via a combination of personally identifying information (including social security number, date of birth, names, and zip code) and encrypted for security. Medical data files were uploaded to a secure cloud server, with their access managed by the blockchain logic.

Unlike other blockchain designs, Gropper's "HIE of One" system~\cite{gropper2016} focused on the creation and use of blockchain-based identities to credential physicians and address the patient matching challenge facing health IT systems. Patients are expected to install a digital wallet on their personal devices to create their blockchain-based IDs, which can then be used to communicate with the rest of the network. Instead of storing patient information, Gropper's system would consume only the blockchain-based ID and use it to secure and manage access to patient data located in EHR systems. 

\subsection{Differentiating Our Research Focus of FHIRChain from Related Work}
This paper presents our blockchain-based framework, called FHIRChain, whose architectural choices were explicitly designed to meet key technical requirements defined by the ONC interoperability roadmap. Our design differs from related work on blockchain infrastructures and associated consensus mechanisms since it is decoupled from any particular blockchain framework and instead focuses on design decisions of smart contract and other blockchain-interfacing components. FHIRChain is thus compatible with any existing blockchains that support the execution of smart contracts. 

In the remainder of this paper we describe how our FHIRChain-based DApp demonstrates the use of digital health identities that do not directly encode private information and can thus be replaced for lost or stolen identities, even in a blockchain system. While our approach is similar to the use of digital IDs in the \textit{HIE of One}\cite{gropper2016} system, FHIRChain provides a more streamlined solution. In addition, we incorporate a token-based access exchange mechanism in FHIRChain that conforms with the FHIR clinical data standards.  Finally, we leverage public key cryptography to simplify secure authentication and permission authorizations, while simultaneously preventing attackers from obtaining unauthorized data access. 

\section{Technical Requirements for Blockchain-Based Clinical Data Sharing}
\label{requirements}
The ``Shared Nationwide Interoperability Roadmap" defines technical requirements and guiding principles for creating interoperable health IT systems~\cite{desalvo2015connecting}. Based on our experiences to date, we contend that crafting a blockchain architecture to meet these requirements necessitates overcoming significant challenges to utilize blockchain technology in healthcare most effectively. 

This section first analyzes five key technical requirements fundamental to clinical data sharing systems and then discusses the implications of these requirements on blockchain-based architectures.  Sections~\ref{fhirchain} and~\ref{results} subsequently describe how we developed and applied our FHIRChain blockchain-based architecture to create a decentralized app (DApp) that meets the ONC requirements in the context of collaborative clinical decision making.

\subsection{Requirement 1: Verifying Identity and Authenticating All Participants}
\paragraph{ONC requirement summary} 
The ONC requirements state that an identity ecosystem should be employed to minimize identity theft and provide redress in case of medical identity fraud, while complying with individual privacy regulations. Providers, hospitals, and their health IT systems should be easily identity-proofed and authenticated when exchanging electronic health information.  Healthcare systems today, however, lack ``consistently applied methods and criteria" for identity proofing and authentication across organizations~\cite{desalvo2015connecting}.  For example, different network service providers have different policies or requirements and may not acknowledge the methods applied by other network service providers.  

One of the most popular---and least complex---approaches to exchange data is through direct secure messaging~\cite{desalvo2015connecting}.  For example, the Direct project~\cite{directhealthit} was launched to create a standard way for participants to send authenticated, encrypted health information directly to known, trusted recipients over the Internet.  Providers or care centers using EHR systems \textit{without} Direct integration, however, cannot benefit from the direct exchange capability.

\paragraph{Implications for blockchain-based system design} For a blockchain-based system, storing identification information (such as personal email) directly on-chain is problematic~\cite{greenspanDB}. In particular, a property of blockchains is information ``openness,'' \textit{i.e.}, all data and associated modification records are immutably recorded and publicly available to all network participants. In the case of Bitcoin, data is open to everyone with Internet access~\cite{nakamoto2008bitcoin}, whereas in a non-public blockchain (such as a consortium blockchain~\cite{buterin2013ethereum}) data access is limited only to authenticated blockchain participants. 

To meet the requirement of openness while complying to health privacy regulations~\cite{centers2003hipaa}, a blockchain-based system should thus support user identity-proofing and authentication while encapsulating sensitive personal information.  Section~\ref{sec:req1} shows how FHIRChain addresses this identifiability and authorization requirement via digital health identities based on public key cryptography~\cite{menezes1996handbook}.
 
\subsection{Requirement 2: Storing and Exchanging Data Securely }
\label{req2}
\paragraph{ONC requirement summary} 
The ONC requirements state that data should be shared securely and privately without unauthorized or unintended alteration, while making the information available to authorized parties. Data encryption is a recommended both when data is sent over networks (data-in-motion) and when it is stored (data-at-rest). Management and distribution of encryption keys must be "secure and tightly controlled"~\cite{desalvo2015connecting}. 

\paragraph{Implications for blockchain-based system design} There has been recent interest~\cite{al2017medibchain,yue2016healthcare} in using blockchain technologies as decentralized storage for encrypted health data. As discussed in Section~\ref{background}, however, the open and transparent nature of blockchain raises privacy concerns when attempting to integrate blockchain into the health IT domain. Although sensitive data can be encrypted, flaws in encryption algorithms or software implementations may expose the data contents in the future. To ensure long-term data security, therefore, a data storage design should be ``simple'' to minimize software bugs~\cite{sheaSimple}, \textit{e.g.}, by not storing sensitive data (encrypted or not) on-chain, yet still enable data flow from one user to another~\cite{zhang2017metric}. 

Another implication of storing data on a blockchain is scalability. All blockchain transactions (such as storing data in a smart contract and modifying the data) and data records are distributed as an entire copy to all blockchain nodes. In a public blockchain, moreover, transaction fees are paid to miners to reward their validation efforts , as described in Section~\ref{background}. As new data is added or modified, each change must be propagated to all nodes, raising scalability challenges and potentially incurring significant long-term operational costs. Section~\ref{sec:req2} shows how FHIRChain addresses this requirement via a hybrid on-chain/off-chain storage model.
 
\subsection{Requirement 3: Consistent Permissioned Access to Data Sources}
\label{req3}
\paragraph{ONC requirement summary} 
The ONC advocates ``computable privacy'' that represents and communicates the permission to share and use identifiable health information~\cite{desalvo2015connecting}. Individuals should be able to document their permissions electronically, which are then honored as needed. Permission authorizations to receive or access an individual's clinical data should be accurate and trustworthy, requiring both the data requestor and holder to have a common understanding of what is authorized. 

\paragraph{Implications for blockchain-based system design} Unfortunately, smart contract operations only occur in the blockchain space to ensure deterministic outcomes. Services (such as OAuth~\cite{hardt2012oauth}) that exist off the blockchain therefore cannot be used. Given this constraint, incorporating other alternatives to provide data access permissioning should be a key component of a blockchain-based design. Section~\ref{sec:req3} shows how FHIRChain addresses this requirement via a token-based permission model.

\subsection{Requirement 4: Applying Consistent Data Formats}
\paragraph{ONC requirement summary} 
To satisfy interoperability needs, the ONC requirements state that health IT systems should be implemented with an ``intentional movement and bias''~\cite{desalvo2015connecting} toward a clinical data standard identified by ONC’s recently finalized \textit{Interoperability Standards Advisory}~\cite{isa}. The data exchanged should be structured, standardized,
and contain discrete (granular~\cite{kim2013structured}) information. Likewise, standards should use metadata to communicate their context along with pieces of structured data. 

\paragraph{Implications for blockchain-based system design}  To provide collaborative clinical decision support, health IT systems must present shared data to clinicians in a structured and readable format~\cite{kawamoto2013key}.  This requirement implies the enforcement of existing, commonly accepted clinical data standard(s), rather than introducing new data exchange formats.  Section~\ref{sec:req4} shows how FHIRChain addresses this requirement by enforcing the FHIR standard.

\subsection{Requirement 5: Maintaining Modularity}
\paragraph{ONC requirement summary} 
The ONC requirements state that since technology inevitably changes over time, health IT system designs should be capable of evolving by maintaining modularity. When divided into connected, modular components, health IT systems become more resilient to change with increased flexibility. In turn, these properties enable the adoption of newer, more efficient technologies over time without rebuilding the entire system. 

\paragraph{Implications for blockchain-based system design} Modularity requires a carefully crafted design to avoid ``information lock-in'' due to the immutability of smart contracts.  Every change to a smart contract code creates a new contract instance on the blockchain, nullifying previous versions and their data.  To minimize dependencies and the need to upgrade, therefore, smart contracts should be loosely coupled with other components in the system. Section~\ref{sec:req5} shows how FHIRChain addresses this requirement by applying the \textit{model-view-controller} (MVC) pattern~\cite{leff2001web}.

\section{FHIRChain: a Blockchain-Based Architecture for Clinical Data Sharing}
\label{fhirchain}
This section first presents an overview of FHIRChain, which is a blockchain-based architecture we designed to meet the ONC requirements for secure and scalable sharing of clinical data described in Section~\ref{requirements}.  We then explain why we made specific architectural decisions in FHIRChain to address each requirement and how they solve the five challenges facing blockchain technology described in Section~\ref{requirements}.  

\subsection{FHIRChain Overview}
\begin{figure}[htpb]
\centering
\includegraphics[width=0.8\columnwidth]{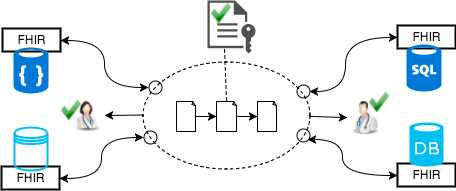}
\caption{Architectural Components in FHIRChain}
\label{fig:fhirchain1}
\end{figure}

Figure~\ref{fig:fhirchain1} shows the FHIRChain architecture we devised to address  key ONC technical requirements. 
This architecture provides a general data sharing solution applicable to a wide range of health IT systems. It also serves as the basis for our decentralized app (DApp) prototype describe in Section~\ref{results}, which customizes FHIRChain to support collaborative clinical decision making using a case study of cancer care in telemedicine.

The dashed ellipse in Figure~\ref{fig:fhirchain1} represents a blockchain component that mediates data sharing between collaborating medical professionals (represented by providers with green check marks).  Clinical data silos are represented by heterogeneous database symbols, which we normalized with the FHIR standards to enforce a common structure of shared data.  Secure database connectors (represented as small circles) connect siloed data sources to the blockchain by exposing secure access tokens to data references that can be obtained only by authorized entities.  The secure tokens are recorded in a smart contract (represented by linked documents) for decentralized access and also traceability.  

In addition to storing secure access tokens, the smart contract also maintains an immutable timestamped transaction log (represented as a keyed file symbol) of all events related to exchanging and actually consuming these tokens.  These logs include specific information regarding what access has been granted to which user by whom, who has consumed which token to access what resource, etc.  To ensure the validity of shared data, FHIRChain can be configured to only approve participation from certified clinicians and healthcare organizations with a membership registry.

\subsection{FHIRChain Architectural Decisions that Address Key ONC Technical Requirements}

Below we explain why specific architectural decisions were made to address each ONC requirement presented in Section~\ref{requirements}.

\subsubsection{Addressing Requirement 1:  Verifying Identity and Authenticating All Participants}
\label{sec:req1}
\paragraph{Context} Blockchains like Ethereum and Bitcoin provide pseudo-anonymous personal accounts (\textit{i.e.}, public addresses composed of random hash values) for users to transact cryptocurrencies. These native identities, however, do not address healthcare requirement for identifiability or authentication of all participants.  

\paragraph{Problem} By design, public blockchains are globally accessible to anyone with Internet access and allow users to hold any number of blockchain accounts to minimize the identifiability of account holders.  This ONC requirement, however, specifies that all U.S. healthcare participants should be identifiable, implying the need for an entirely separate, traceable user base from blockchains' native identities.  A key problem is thus how to properly define identities for healthcare users participating in clinical data sharing, while protecting sensitive personal information on the blockchain. 

\paragraph{Design choice $\rightarrow$ use of a digital health identity} Inspired by the success of secure shell (SSH)~\cite{ylonen2006secure} and blockchain address generation mechanism, FHIRChain employs public key cryptography~\cite{menezes1996handbook} to create and manage health identities.  In public key cryptography, a pair of mathematically related public and private keys is used to create digital signatures and encrypt data.  Since it is computationally infeasible to obtain the private key given its paired public key, these public keys can be shared freely, thereby allowing users to encrypt content and verify digital signatures. In contrast, private keys are kept secret to ensure only their owners can decrypt content and create digital signatures.  

FHIRChain generates a cryptographic public/private key pair (also used for encryption, as described in Section~\ref{sec:req3}) for each participating provider, \textit{e.g.}, in-house providers and remote physicians in telemedicine clinics. The public keys represent users' digital health identities. These identities are recorded in the blockchain for both identity- and tamper-proofing, thereby ensuring that users holding the corresponding private keys can be authenticated to use FHIRChain's data sharing service.

FHIRChain's design applies a smart contract to maintain health users' identifiability without exposing personal information on the blockchain.  It also replaces the need for a traditional username/password authentication scheme with the use of a public/private cryptographic key pair for authentication.  In a general clinical setting, these digital health identities (\textit{i.e.}, their private keys) would be hard to manage for patients.  FHIRChain, however, only creates these identities for clinicians to facilitate data sharing, which enables more effective collaborative decision making for patients.

\subsubsection{Addressing Requirement 2: Storing and Exchanging Data Securely}
\label{sec:req2}
\paragraph{Context} A key capability offered by blockchains is their support for ``trustless” transactions between parties who lack trust relationships established between them. Bitcoin is the most common example of this “trustless” exchange via its native cryptocurrency.  Blockchains are peer-to-peer by nature and thus contribute to the ubiquitousness of digital assets being transacted.

\paragraph{Problem} Health data represented via digital assets are more complex and harder to share \textit{en masse}. There are also privacy and security concerns associated with its storage in an ``open'' peer-to-peer system (\textit{i.e.}, public blockchains), such as encryption algorithms applied to protect data being decryptable in the future~\cite{zhang2017metric}. A key problem is thus how to design a blockchain-based health IT system so that it balances the need for ubiquitous store and exchange and the concerns regarding privacy of the data and scalability of the system. 

\paragraph{Design choice $\rightarrow$ keeping sensitive data off-chain and exchanging reference pointers on-chain} Rather than storing encrypted health data in the blockchain, a more scalable and secure alternative is to store and exchange encrypted metadata referencing protected data (\textit{i.e.}, a reference pointer to a data set), which can be combined with an expiration configuration for short-term data sharing.  Exchanging encrypted reference pointers allows providers to maintain their data ownership and choose to share data at will.  This technique also prevents an attacker who intercepts the encrypted pointers from obtaining unauthorized data access.

FHIRChain attaches a secure connector to each database, as shown in Figure~\ref{fig:fhirchain1}.  Each connector generates appropriate reference pointers that grant access to the data. These reference pointers are digital health assets that can be transacted ubiquitously with reduced risks of exposing the data. 

An added benefit of exchanging metadata \textit{en masse} is more scalability compared to exchanging the original data source.  As discussed in Section~\ref{req2}, each transaction or operation on the blockchain (\textit{e.g.}, querying a smart contract state variable value or updating it) is associated with a small fee paid to the miner for verification and then included onto the blockchain.  Transacting these lightweight reference pointers is more efficient in terms of time and cost in production because small changes to data generally require no modifications to reference pointers.  


\subsubsection{Addressing Requirement 3: Permission to Access Data Sources}
\label{sec:req3}
\paragraph{Context} Data references can be stored on the blockchain for ubiquitous access via a smart contract.  Access rights, however, must be granted only to authorized providers for viewing the data. As discussed in Section~\ref{req3}, OAuth is a popular platform for communicating permissions in web-based apps that are not based on blockchain. 

\paragraph{Problem} Smart contracts cannot directly use external services like OAuth since they do not produce deterministic outcomes that can be verified by blockchain miners. A key problem is thus how to design a mechanism that balances the need of permission authorization for clinical data and blockchain requirements for deterministic outcomes.

\paragraph{Design choice $\rightarrow$ token-based permission model} To overcome the limitation with public blockchains, FHIRChain protects the shared content via a secure cryptographic mechanism called ``sign then encrypt''~\cite{krawczyk2001order}. This design employs the users' digital health identities to encrypt content so that only users holding the correct digital identity private keys can decrypt the content.  FHIRChain also generates a new pair of signing keys for each participant and registers the public portion of signing keys alongside users' digital identities.  

\begin{figure}[hptb]
\centering
\includegraphics[width=0.9\textwidth]{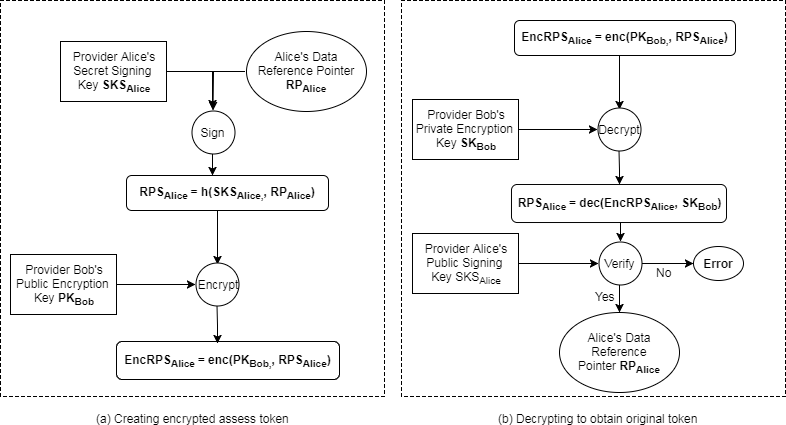}
\caption{Example of the Creation and Retrieval of an Access Token Using FHIRChain. }
\label{fig:alice}
\end{figure}

To concretely demonstrate this workflow, Figure~\ref{fig:alice} provides an example of using FHIRChain to create and retrieve an access token.
Suppose provider \textit{Alice} would like to initiate sharing of her patient's data, denoted as \textit{D\textsubscript{Alice}} (with a reference pointer, denoted as \textit{RP\textsubscript{Alice}}) with another provider \textit{Bob}.  FHIRChain creates a digital signature on the shared content \textit{RP\textsubscript{Alice}}, with \textit{Alice}'s private signing key \textit{SKS\textsubscript{Alice}} for tamper-proofing as a first step. With \textit{Bob}'s public encryption key, \textit{PK\textsubscript{Bob}}, FHIRChain encrypts the signed \textit{RPS\textsubscript{Alice}} to obtain an encrypted token \textit{EncRPS\textsubscript{Alice}}, and then stores \textit{EncRPS\textsubscript{Alice}} in a smart contract for ubiquitous access.  
 
When \textit{Bob} wants to obtain the content \textit{Alice} sent, he must use his corresponding private encryption key \textit{SK\textsubscript{Bob}} to decipher the real content of \textit{EncRPS\textsubscript{Alice}}. \textit{Bob} also verifies that this content was indeed provided by \textit{Alice} with her public signing key \textit{PKS\textsubscript{Alice}}.  This authentication process is automated by the DApp server component interfacing the smart contract, as discussed in Section~\ref{sec:req5}.

Digital signing ensures that a resource is indeed shared by the sender and is not tampered with. Likewise, encryption protects the information against unauthorized access and spoofing. The data requestor's access to a resource can be approved or revoked at any time via a state update in the smart contract by the data holder where all permissions are logged. 

Role-based or attribute-based permissions can also be implemented off-chain in the same manner as in a traditional centralized system (\textit{e.g.}, via Active Directory). In this case, a meta-cryptographic key pair would be created for each role or type of attribute and securely stored within the system’s database. The system can then be configured so that only allows users meeting certain permission criteria to use the key for data access, while shielding users from unessential details.

\subsubsection{Addressing Requirement 4: Consistent Data Formats}
\label{sec:req4}
\paragraph{Context} Clinical research data can exist in various formats and structures, which may or may not be meaningful when shared with other providers from different organizations.

\paragraph{Problem} 
Blockchain-based health IT systems should facilitate data sharing, while adhering to some existing standard(s) for representing the clinical data. A key problem is thus how to design a blockchain-based architecture to enforce the application of existing clinical data standard(s).

\paragraph{Design choice $\rightarrow$ enforcing FHIR standards} HL7's FHIR standards use JSON~\cite{crockford2006application}, which is a popular format for exchanging clinical information. JSON is more compact and readable compared to the XML format used by other data formatting standards, thereby enabling more efficient transmission of JSON-encoded data. It is also compatible with many software libraries and packages. As more health IT systems upgrade their data exchange protocols to comply to FHIR standards, FHIRChain enforces the use of FHIR to shared clinical data by validating whether the generated reference pointers follow the FHIR API standards~\cite{bender2013hl7}.

\subsubsection{Addressing Requirement 5: Maintaining Modularity}
\label{sec:req5}
\paragraph{Context} Health IT system updates and/or upgrades are necessary to adopt more efficient, secure, or prevalent technology as it advances. 

\paragraph{Problem} 
If functions in a smart contract have too many dependencies on the rest of a health IT system, then each upgrade to the system must deploy a new contract, which requires restoring data from previous versions to prevent loss. A key problem is thus how to design a modular data sharing system that minimizes the need to create new versions of existing contracts when the system is upgraded. For example, when more user friendly features are needed, a good design should separate those updates from the underlying back-end services so that a change in the user interface does not require modifications of the server or blockchain component. 

\paragraph{Design choice $\rightarrow$ applying the model-view-controller (MVC) pattern} The \textit{MVC} pattern~\cite{leff2001web} separates a system into three components: (1) the \textit{model}, which manages the behavior and data of a system and responds to requests for information about its state and instructions to change state, (2) the \textit{view}, which manages the display of information, and (3) the \textit{controller}, which interprets user inputs into appropriate messages to pass onto the \textit{view} or \textit{model}. 

The FHIRChain architecture applies the \textit{MVC} pattern to separate concerns with individually testable modules as follows: (1) a model in the form of an immutable \textit{blockchain component} is used to store necessary meta data via smart contracts; (2) a view provides a front-end \textit{user interface} that accepts user inputs and presents data; (3) a controller is a \textit{server} component with control logic that facilitates interactions with data between the \textit{user interface} and \textit{blockchain component}, such as queries, updates, encrypting and decrypting contents; and (4) a controller-invoked \textit{data connector} service is used to validate the implementation of FHIR standards and create reference pointers for the data sources upon requests from the server.  

\begin{figure}[hptb]
\centering
\includegraphics[width=0.9\textwidth]{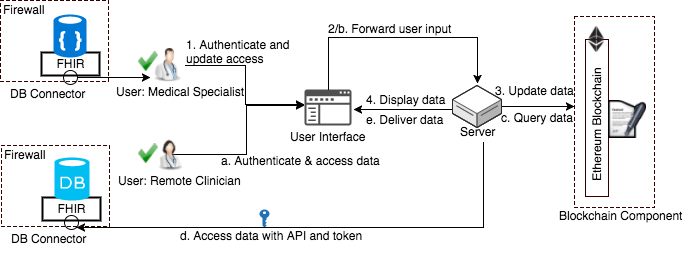}
\caption{Composition and Structure of the FHIRChain Architecture with Modular Components. }
\label{fig:fc2}
\end{figure}
The workflow for updating data access is shown in Figure~\ref{fig:fc2} by the following steps 1-4: 
\begin{enumerate}
\item A user first authenticates through the user interface (UI), and when successfully authenticated, data access permission request can be input to the system;
\item The UI forwards user's request to the server; 
\item The server logs permissioned or revoked access in the blockchain component (BC); and 
\item The server updates UI with proper response to notify the user. 
\end{enumerate}

Likewise, the workflow for accessing a data source is outlined in the following steps a-e: 
\renewcommand{\labelenumi}{\alph{enumi})}
\begin{enumerate}
\item The user first authenticates via the UI, and when successfully authenticated data access request can be input to the system; 
\item UI forwards user’s request to the server; 
\item The server queries BC for current user's access token(s); 
\item When permission is valid, the server decodes the access token(s) with correct keys supplied by user and uses the decrypted reference pointer to obtain actual data from the DB connector to the proper database; 
\item When data has been retrieved from the data source via DB connector, the server updates UI to display data in a readable format. 
\end{enumerate}

FHIRChain stores all relevant information in smart contracts, decoupling data store from the rest of the system. This decoupling enables future upgrades to all other components without losing access to---or locking out---existing users or their permission information.

\section{Case Study: Applying FHIRChain to Create a Prototype DApp}
\label{results}

This section first describes the structure and functionality of a \textit{decentralized app} (DApp) that customizes the FHIRChain architecture described in Section~\ref{fhirchain} to support collaborative clinical decision making via a remote tumor board case study.  We then analyze the benefits and limitations of our DApp case study.

\subsection{Overview of the FHIRChain DApp Case Study}
The FHIRChain DApp is written in Javascript. It consists of $\sim$1,000 lines of core app code that interacts with a private testnet of the Ethereum blockchain and three Solidity smart contracts, each containing $\sim$50 lines of code.  Our DApp customizes the FHIRChain architecture in a private Ethereum testnet to address the various ONC requirements described in Section~\ref{requirements}.

This DApp has an intuitive user interfacing portal that facilitates the sharing and viewing of patient cancer data for a remote tumor board to collaboratively create treatment plan for cancer patients.  In addition, the DApp implements a notification service~\cite{zhang2017applying} that broadcasts events to appropriate event subscribers.  The FHIRChain DApp notification service is used to alert collaborative tumor board members when new data access is available for review.   

\begin{figure}[t]
\centering
\includegraphics[width=0.7\textwidth]{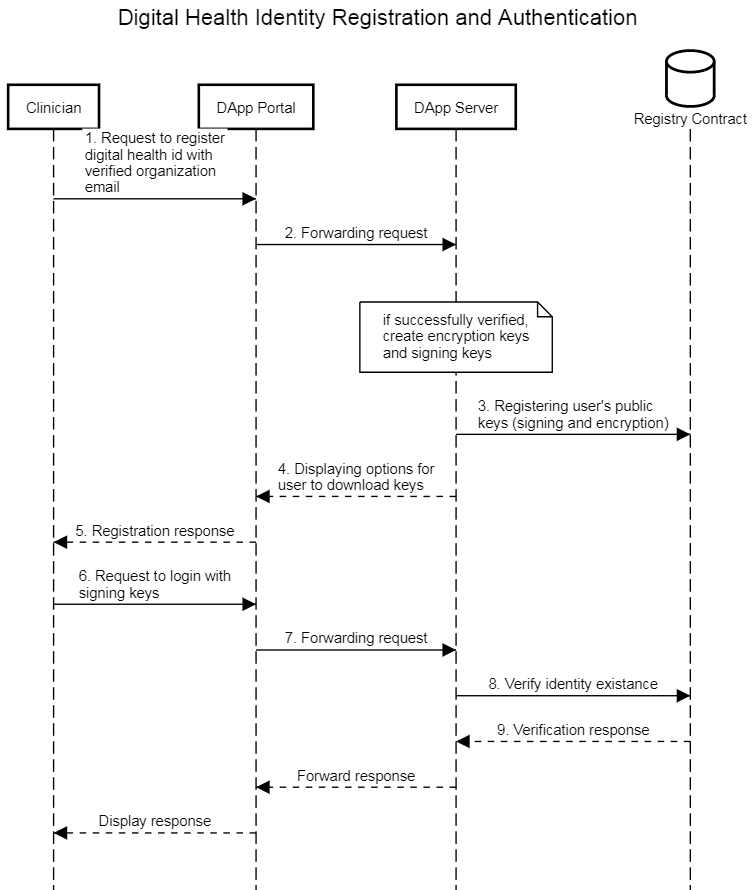}
\caption{Workflow of the User Registration and Authentication Process in the FHIRChain DApp. }
\label{fig:reg}
\end{figure}

\textbf{Verifying identity and authenticating participants with digital identities, as discussed in Section~\ref{sec:req1}}. Our DApp contains a \textit{Registry} smart contract that maintains the digital health identities of providers who registered with our app.  The registry maps provider email addresses (or phone numbers) from a public provider directory to both their public encryption (used as digital identity) and signing keys, which are generated automatically at user registration time.  Figure~\ref{fig:reg} demonstrates the user registration and authentication workflow.

\textbf{Storing and exchanging data securely with FHIR-based reference pointers, as discussed in Sections~\ref{sec:req2} and~\ref{sec:req4}}. Our DApp defines two cancer patient databases and referencing paths to patient data entries using the open-source HapiFHIR~\cite{hapi} public test server.  Validation of the FHIR implementation is performed via regular expression parsing of the paths against the FHIR APIs~\cite{bender2013hl7}. 

\begin{figure}[t]
\centering
\includegraphics[width=0.8\textwidth]{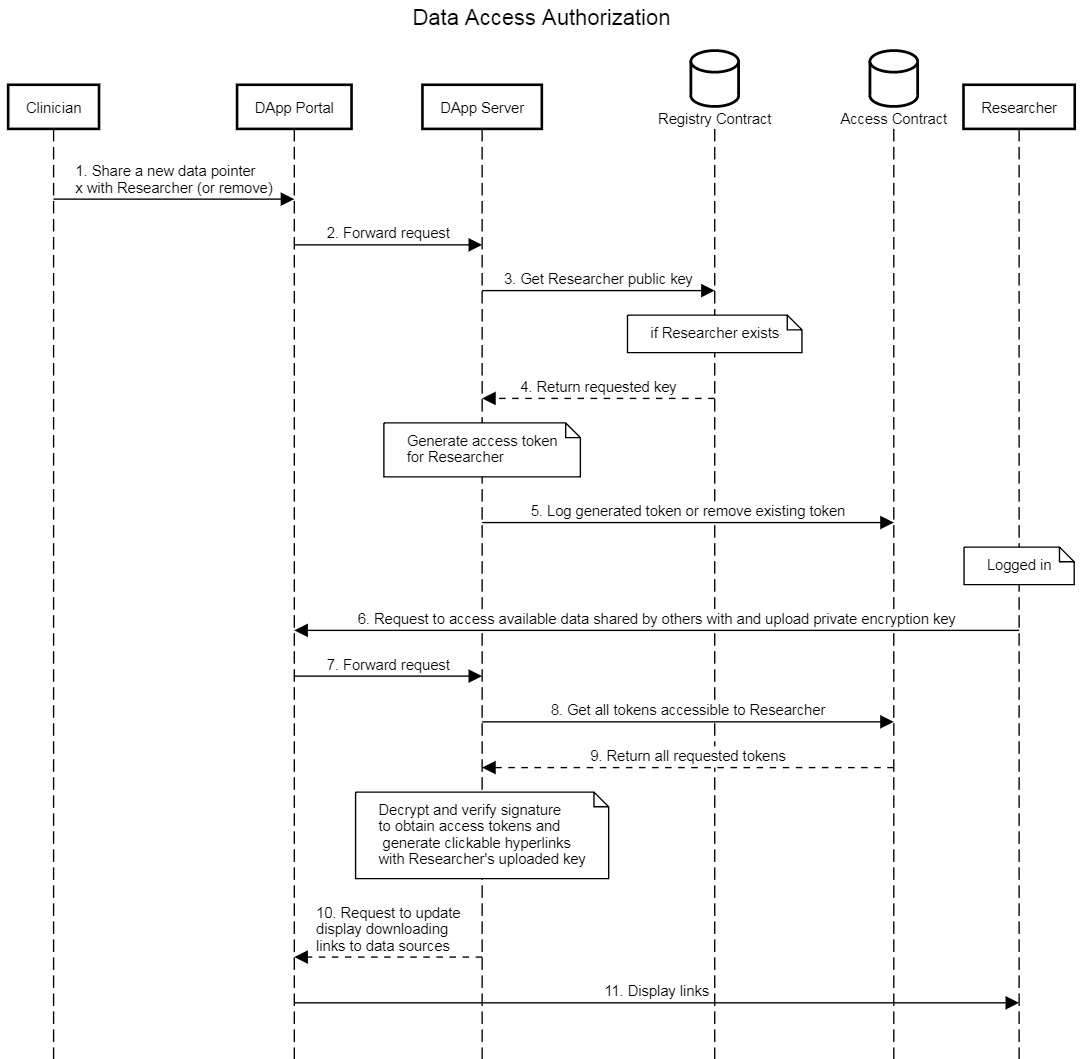}
\caption{Workflow of Access Authorization in the FHIRChain DApp. }
\label{fig:access}
\end{figure}

\textbf{Permissioning data access with token-based exchange, as discussed in Section~\ref{sec:req3}}. Our DApp also contains an \textit{Access} smart contract that logs all user interactions and requests on the portal, \textit{e.g.}, what resource is shared or no longer shared with which provider by whom and when.  These access logs are structured as a mapping between user digital health identities (public encryption keys) and authorizations to custom-named access tokens (represented as a nested object associated with a \textit{true/false} boolean value indicating if an access token access is granted for a provider).  If an access revocation occurs, authorization is set to \textit{false} and the associated token is set to an empty value. The workflow of this process is shown in Figure~\ref{fig:access}.

\begin{figure}[t]
\centering
\includegraphics[width=1.0\textwidth]{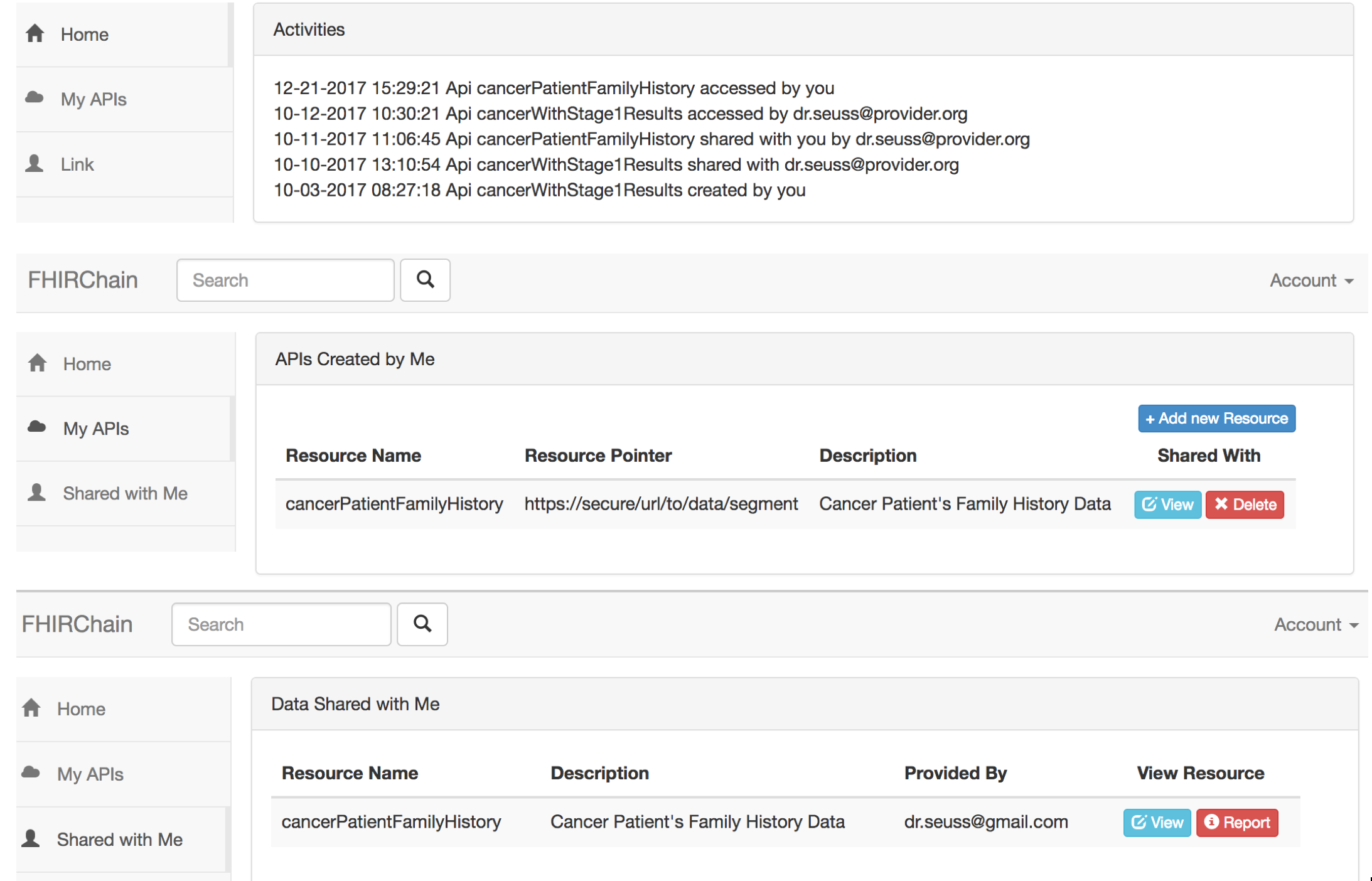}
\caption{Screenshot of Our FHIRChain-based DApp User Interface.}
\label{fig:ui}
\end{figure}

\textbf{Maintaining modularity with the MVC pattern, as discussed in Section~\ref{sec:req5}}. The \textit{view} component is a user interfacing portal that accepts provider user input, including registration and authentication credentials (corresponding keys) and data access information (\textit{e.g.}, tumor board member email to query, a reference pointer to securely access data, and approval/revocation of access).  Figure~\ref{fig:ui} is a screenshot of our DApp, presenting the following features (1) display recent sharing events related to the user, (2) display reference pointer APIs created by logged in user and available actions, and (3) display all references shared with logged in user and the option to view data. 

The portal then forwards the user requests along with data input to the \textit{sever} component, where all the complex logic is encapsulated.  

Our FHIRChain DApp \textit{server} performs all functions and control logic, including verifying provider user email account, generating cryptographic keys, token creation via signing and encryption, token retrieval via decryption and signature verification, forwarding requests and delegating tasks between the \textit{portal} and \textit{blockchain}.  The \textit{blockchain} component is an independent \textit{model} component containing two smart contracts for ubiquitous storing and persisting event logs of data access. 

\subsection{Benefits of Our FHIRChain DApp Case Study}
Our FHIRChain Dapp case study achieved the following benefits:
\begin{itemize}
\item \textbf{Increased modularity}. To increase modularity, we applied the ``separation of concerns'' principle~\cite{ossher2001using} to decompose our DApp into independent components. FHIRChain employs a peer-to-peer API exchange protocol that references data pointers stored in a smart contract on the blockchain. In this design,  exchanged information becomes lightweight, which increases scalability since system performance remains the same regardless of the original size of the data. Likewise, data is not transmitted electronically across institutional boundaries, thereby reducing the risk of data being compromised. 

\item \textbf{Scalable data integrity}. To ensure scalable data integrity, our design maintains a hash of the original data to exchange in addition to the reference pointer of the data.  Suppose that the original data being exchanged is of size \(N\) and that the size of its reference pointer is \(\epsilon\).  The total amount of data stored on-chain in terms of space complexity is then \(O(hash(N) + \epsilon)\).  Since the hashed output of a variable-length input can be a fixed value, it consumes a constant amount of space.  The size of a data reference pointer would be scalably smaller than the actual data size.  This design therefore enhances scalability by using constant-sized representations of the data, rather than using the actual data. 

\item \textbf{Fine-grained access control}. To enable fine-grained access control, permissions to access a data source can be given or revoked at will by providers across various institutions regardless of their trust relationships.  By implementing the FHIR standards, more granular access can be granted to selected pieces of data rather than an entire document, which also increases data readability.  Moreover, all events related to data sharing and data access are logged in a transparent history for auditability.   

\item \textbf{Enhanced trust}. The DApp applies public key cryptography, which enhances trust to participants in the following ways:
\begin{itemize}
\item \textbf{Identifiability and authentication}. Given the computation power today, it is infeasible to impersonate a user without knowing their private key, and the only way a user can be authenticated to use our service is to provide the correct private key paired with their public key registered on the blockchain. On the other hand, it is trivial to create a new public/private key pair in case of a user's private key being lost or stolen. This ``digital identity" approach has been successfully adopted in Estonia’s government and healthcare infrastructure~\cite{alvarez2009internet}. 

\item \textbf{Permission authorization}. With public key encryption securing their data reference pointers, users can trust that none other than the intended data recipient can view what they have shared. FHIRChain never shares the reference pointer with any user. Instead, RP is used to display the data content when it is decrypted with an authorized user's private key. In addition, users can approve or revoke data access at any time, and the request takes effect immediately. 
\end{itemize}
\end{itemize}

\subsection{Limitations of Our FHIRChain DApp Case Study}
Since our FHIRChain DApp was designed based on several assumptions it incurs the following limitations: 
\begin{itemize}

\item \textbf{Does not address semantic interoperability}. FHIRChain cannot address data exchange challenges related to semantic interoperability that are not yet fully captured by the FHIR standards.  To provide semantics to clinical data, therefore, manual inspection and mapping of predefined ontologies from medical and health data experts are required, which remain the focus of our future research in this space. 

\item \textbf{May not be compatible with legacy systems not supporting FHIR.} Many legacy systems may use other messaging standards, such as the more prevalent HL7 v2 standards~\cite{dolin2006hl7}, and do not support FHIR protocols.  The goal of this paper, however, is to present the underlying representations and theories of our blockchain-based system. Although we advocate FHIR in the paper because it has been used quite frequently and it supports fine-grained data exchange, the principles behind the system described here can also be used with other standards like HL7 v2~\cite{dolin2006hl7}.

\item \textbf{Cannot control clinical malpractice}. The intended users of FHIRChain are clinicians interested in collaboratively providing clinical decision support for remote patients.  Our current design trusts that the data being exchanged using our DApp is not abused, misused, or unethically redistributed by users.  Our future work will explore options to minimize these risks, such as tracking data credibility using cryptographic hashing or zero knowledge proofs~\cite{rackoff1991non} (ability to demonstrate the truth of a statement without revealing additional information beyond what it’s trying to prove~\cite{zeroKnowledge}) along with each reference pointer.  Naturally, clinical malpractice may still occur (as in any other health IT system) since we cannot fully control these human behaviors. 

\item \textbf{DApp deployment costs}. Unlike existing public blockchain, such as Ethereum, our DApp is developed using a private testnet that imposes no interaction costs (\textit{e.g.}, transaction fees).  Our DApp would thus not be free of charge if deployed on a public blockchain. The convenience provided by a public blockchain, however, may justify the cost of usage versus the costs of licensing, running, and maintaining a private clinical data exchange infrastructure.  
\end{itemize}

To overcome these limitations in future work, we will deploy our DApp in a permissioned consortium blockchain platform with trusted parties to ensure consensus through a variation of proof-of-work that incentivizes mining with cryptocurrency rewards.  For instance, \cite{Ekblaw2016} proposes to use aggregated data as mining rewards in their system, while MultiChain~\cite{greenspan2015multichain} enforces a round-robin mining protocol in their blockchain. With the ability to replace monetary incentives to maintain consensus on the blockchain, the cost to use this blockchain-based service will be lower in the long run, although the initial deployment may still be expensive.

Although permissioned systems may be prone to collusion due to the 51\% attack problem~\cite{buterin2013ethereum}, the permissioned system used for healthcare would be maintained and managed by relatively large-scale entities/stakeholders within the healthcare industry. Unless majority of them (major hospitals, insurance companies, \textit{etc}.) collude, therefore, the chance of experiencing this type of attack is quite low. Moreover, legal actions would most likely occur immediately upon the attack. 
\section{Concluding Remarks}
\label{conclusion}

This paper described the FHIRChain prototype we designed to provide patients with more collaborative clinical decision support using blockchain technology and the FHIR data standards. Complemented by the adoption of public key cryptography, our FHIRChain design addressed five key requirements provided by the ONC interoperability roadmap, including user identifiability and authentication, secure data exchange, permissioned data access, consistent data formats, and system modularity. 

The following are the key lessons we learned from designing and implementing our DApp based on FHIRChain:
\begin{itemize}
\item \textbf{FHIRChain can provide trustless, decentralized storage for necessary meta information and audit logs}.  FHIRChain alleviates proprietary vendor-lock found in conventional health IT systems by leveraging its blockchain component as a decentralized storage of necessary reference information as secure access points into those databases.  It enables the sharing of clinical data without established trusts, providing clinicians with secure and scalable collaborative care decision support. In addition, each public key generated for a user is stored in the blockchain via a smart contract used to associate healthcare participants with their digital identities. Similarly, permission authorizations established between those participants are recorded in a smart contract as well, creating a traceable permission database with an audit log of data exchange history (\textit{i.e.}, meta information involved during the data exchange and not the actual data).  Storing these data on the blockchain ensures that our app is not subject to a single point of failure or corruption of records so that it is always accessible by healthcare participants.

\item \textbf{FHIRChain facilitates data exchange without the need to upload/download data thus maintains data ownership.}  The FHIR standards provide resource APIs to reference specific pieces of structured data while maintaining original data ownership.  By adopting FHIR and combining it with blockchain technologies, FHIRChain creates lightweight reference pointers to siloed databases and exchange these pointers via the blockchain component instead of actual data.  For telemedicine clinics or clinics in rural areas in particular, this approach can overcome network limitations by enabling scalable data sharing without requiring data to be uploaded to some other centralized repository, through which data can be shared and downloaded by other parties.  In addition, this approach reduces risks of compromised data and ensures that original data ownership is respected.  The reference pointers are encrypted with the intended recipient’s public key, \textit{i.e.}, digital identity to permission data access.  When successfully authenticated (\textit{i.e.}, reference pointers are correctly decrypted) the data will be downloaded directly from the source and present properly formatted data to the user. 

\item \textbf{Public key cryptography can be effective for managing digital health identity in data sharing}.  FHIRChain creates public keys as digital health identities associated with each collaborating care entity (provider or organization administrator).  The benefits to this strategy include: (1) \textit{easy authentication} since a clinician only needs to provide their private key associated with their identity, (2) \textit{integrity} since by signing the exchanged reference pointers FHIRChain can easily verify that it was provided by the signed provider and has not been modified, and (3) \textit{remedy to lost or stolen keys} since a new key can be created easily to replace the old key and associate with the same user. There is a drawback, however, to using digital identities for patients in a general clinical setting. Managing these identities---private keys---is hard because private keys are harder to remember than conventional passwords and require technical training for patients to manage their own keys.  Nevertheless, there are approaches for managing private keys for larger populations, such as using key wallets~\cite{even1984electronic,nakamoto2008bitcoin} or embedding private keys to physical medical ID cards~\cite{anthes2015estonia}. 
\end{itemize}
		
In summary, our FHIRChain-based DApp demonstrates the potential of blockchain to foster effective healthcare data sharing while maintaining the security of original data sources.  FHIRChain can be further extended to address other healthcare interoperability issues, such as coordinating other stakeholders (\textit{e.g.}, insurance companies) across the industry and providing patients with easier (and secure) access to their own medical records.  

In our future work, we plan to refine the simulations for more rigorously evaluating the performance of our FHIRChain prototype.  We will do so by deploying and comparing a number of different blockchain configurations in a testbed environment, such as using the blockchain template provided by Amazon Web Services~\cite{awsBlockchain}.  Moreover, we will research techniques for identity management targeting the patient population.


\section*{Acknowledgements}
The work presented in this paper was sponsored in part by funding from Varian Medical Systems.


\section*{References}
{\small
\bibliographystyle{model1-num-names}
\bibliography{main.bib}
}







\end{document}